\documentstyle[osa,epsfig]{revtex}
\begin{document}

\title{Polymer nano-droplets forming liquid bridges in chemically
structured slit pores: A computer simulation}

\author{Jacqueline Yaneva$^1$, Andrey Milchev$^{1,2}$, and Kurt Binder$^2$}
\maketitle
\medskip

$^1$ Institute for Physical Chemistry, Bulgarian Academy of
Sciences, 1113 Sofia, Bulgaria

\medskip

$^2$ Institut f\"ur Physik, Johannes-Gutenberg-Universit\"at
Mainz, 55099 Mainz, Germany

\bigskip \abstract{\small Using a coarse-grained bead-spring model
of flexible polymer chains, the structure of a polymeric
nanodroplet adsorbed on a chemically decorated flat wall is investigated
by means of Molecular Dynamics simulation. 
We consider sessile drops on a lyophilic (attractive for the monomers) region of 
circular shape with radius $R_D$ while the remaining part of the 
substrate is lyophobic. The variation of the droplet shape, including 
its contact angle, with $R_D$ is studied, and the density profiles 
across these droplets also are obtained.

In addition, the interaction of droplets adsorbed on two walls
forming a slit pore with two lyophilic circular regions just
opposite of one another is investigated, paying attention to the
formation of a {\em liquid bridge} between both walls. A central result
of our study is the measurement of the force between the two substrate 
walls at varying wall separation as well as the
kinetics of droplet merging. Our results are compared to various 
phenomenological theories developed for liquid droplets of mesoscopic 
rather than nanoscopic size.}
\medskip

\section{Introduction}
In the context of the emerging technology of microscale processing
 the use of surface substrates with micropatterns has become
increasingly important for microfluidics applications. Chemically
structured surfaces that exhibit lateral patterns of varying
wettability can be produced by techniques such as photolithography
\cite{1,2}, microcontact printing \cite{3,4,5}, vapor deposition
through grids \cite{6}, domain formation in Langmuir-Blodgett
monolayers \cite{7,8}, electrophoretic colloid assembly \cite{9},
lithography with colloid monolayers \cite{10}, microphase
separation in diblock copolymer films \cite{11}, etc. For
patterned surfaces in the micrometer range, on which liquid
droplets or thin liquid films are adsorbed, fascinating wetting
morphologies have been predicted \cite{12,13,14,15,16,17,18,18a}
(including ``morphological wetting transitions''\cite{12,16,17})
and observed \cite{6}. Liquid bridges between droplets at walls
are also of interest in the context of forces between droplets (or
bubbles, respectively) \cite{19,20}, long range forces between
colloidal particles \cite{21,22}, etc. The theoretical treatments
mostly apply phenomenological quasi-macroscopic concepts (in terms
of interfacial tensions, contact angles, \cite{23,24} etc.); but
for droplets on the micrometer scale the lack of knowledge on the
line tension \cite{24,25,26,27,28,29,30,31,32,33,34,35} is a
serious drawback already, and the understanding of the density
distribution of the droplet near the contact line is a difficult
problem \cite{33,34}. The present drive towards nanoscale
technology creates also more interest in droplets on the nanometer
scale, and in fact some of the techniques mentioned above are well
suited to create surface patterns on the nanoscale \cite{10,11}.
Although sometimes macroscopic concepts do allow reasonable
predictions down to the nanoscale \cite{36,37,38,39}, there is no
guarantee that the phenomenological theories are quantitatively
accurate for bridge formation between polymer nanodroplets. We
study this problem here by Molecular-Dynamics simulation, to
complement existing knowledge on the problem by insight on a
molecular level.

In the next section we shall present our model, which has been
used successfully in previous work to study static and dynamic
properties of polymers in the bulk and at surfaces
\cite{36,37,38,40,41,42,43,44,45,46}. In section 3 we shall
present our simulation results on droplets on single walls, as
well as on the structure of liquid bridges in slit pores of
varying width. Section 4 is devoted to a discussion of dynamic
aspects of bridge formation, while Section 5 summarizes some
conclusions.

\section{Some comments on the model and the simulation technique}
We employ a coarse-grained off-lattice model of polymer chains,
where each chain consists of $N = 32$ effective monomers, which
are connected by anharmonic springs. These effective bonds are
thought to represent groups of a few successive chemical monomers
along the chain, and therefore inclusion of torsional potentials
and even bond-bending potentials is not considered. The springs
are described by the finitely extensible nonlinear elastic (FENE)
potential

\begin{equation}\label{eq1}
U_{FENE}(\ell) = - \frac K 2 R^2\ell n
[1-\frac{(\ell-\ell_0)^2}{R^2} ] \;,
\end{equation}
$\ell$ being the bond length, which can vary in between
$\ell_{\textrm{min}}$ and $\ell_{\textrm{max}}$, and thus the
equilibrium value $\ell _0$ for which $U_{FENE}(\ell _0)=0$, and
$R= \ell_{\textrm{max}}- \ell_0 = \ell_0 - \ell _{\textrm{min}}$.
The spring constant $K$ is taken as in previous work $K/k_BT=40$,
and we again choose $\ell_{\textrm{max}}=1$ as our unit of length,
with $R=0.3$ (hence $\ell_0 =0.7$, $\ell_{\textrm{min}}=0.4$)
\cite{40,41,42,43,44,45,46}.

Between the effective monomers a Morse potential acts,

\begin{equation}\label{eq2}
U_M(r)= \epsilon _M {\exp[-2\alpha (r-r_{\textrm{min}})] - 2
\exp[-\alpha(r-r_{\textrm{min}})]}\;,
\end{equation}
$r$ being the distance between the beads, and the parameters are
chosen as $r_{\textrm{min}}=0.8, \; \epsilon_M=1$ and $ \alpha =
24$. Owing to the large value of $\alpha$, $U_M(r)$ decays to zero
very rapidly for $r >r_{\textrm{min}}$, and is completely
negligible for $r
>1$. This choice of parameters is useful, particularly for
Monte Carlo simulations, since it allows the use of a very
efficient link-cell algorithm \cite{40}. The Theta-temperature for
this model is \cite{42} $k_B\Theta \approx 0.62$. We hence choose
in the following a temperature $k_BT=0.49$, where at zero pressure
the system already is in the state of a dense melt, and to a very
good approximation the gas density is zero for $N=32$. Thus from a
polymer droplet at a surface no chains evaporate into the
surrounding gas.

The adsorbing walls are treated as perfectly flat and
structureless; in particular, no atomic corrugation of the walls
is considered. The interactions between the effective monomers and
the walls are represented by a Lennard-Jones potential, integrated
over a (semi-infinite) substrate \cite{38}

\begin{equation}\label{eq3}
U_{\textrm{wall}}(z)=4\pi \epsilon _{\textrm{w}} [\frac{1}{45}
(\frac{\sigma_{\textrm{wall}}}{z})^9-\frac 1 6
(\frac{\sigma_{\textrm{wall}}}{z})^3]\;,
\end{equation}
where we choose parameters $\sigma_{\textrm{wall}}=1$,
$\epsilon_{\textrm{w}}=0.20$ in the lyophilic part of the
substrate, while in the lyophobic area
$\epsilon_{\textrm{w}}=0.05$ is chosen. Typically, the
lyophilic area is a circle of radius $R_D$, with 3 $\leq R_D \leq 15$.
The total lateral linear dimensions of the simulation box is
chosen $64 \times 64$, so that (for a total number of
${\mathcal{N}}=128$ chains in the polymer droplet) it never can
happen that a droplet interacts with its images generated by the
periodic boundary condition. For a study of single droplets, the
linear dimensions in the perpendicular direction is $L=32$, and
the top wall of the box is taken purely repulsive (omitting the
attractive term from Eq.(\ref{eq3}), and choosing also
$\epsilon_{\textrm{w}}=0.20$). For our study of slit pores, we
choose $5 \leq L \leq 22$, and the potentials from both walls are
chosen exactly of the same type.

The initial preparation of droplets in equilibrium is done by
Monte Carlo methods, applying the Velocity-Verlet algorithm and
keeping the temperature constant by the Nos\'e-Hoover thermostat
\cite{47}. In the case of bridging droplets, however, we resort
to a Langevin thermostat\cite{47a} which provides greater stability of the
algorithm for small separation between the solid planes. For more 
details on our Molecular Dynamics algorithm,
we refer the reader to \cite{38}. Here we only note that due to
the steep variation of the Morse potential a very small
integration time step needs to be used, $\delta t=0.0009$ MD time
units, and typically runs over 1.1 million time steps were carried
out. 

It is certainly useful to translate the time units [t.u.], used in
our simulation, into seconds by mapping our model results to laboratory data.
A typical substance used in numerous experiments on wetting is,
for example, the PDMS (polydimethylsiloxane) melt $(C_2 H_6 O Si)_n$.
In ellipsometric experiments on droplet spreading\cite{Voue} one has
measured a diffusion coefficient ${\cal D}_{diff} \approx 3.3\times
10^{-10} m^2/s$ for chain lengths $N=10$, and ${\cal D}_{diff} \approx 
0.3\times 10^{-10} m^2/s$ for $N=20$. These data can be compared
to our measurements of precursor diffusion\cite{38} in spreading
droplets which yield ${\cal D}_{diff}^{sim} \approx 0.11 \ell_0^2
/ t.u.$ in a melt of chains with $N=32$. Bearing in mind that the bond 
length in PDMS is\cite{Arbe} $\approx 1.59 \AA$, and assuming that, say, 
three chemical units form a persistent length of $\approx 4 \AA$, one 
obtains for the simulation time unit $1 t.u. \approx 0.5 ns$ as a 
rough estimate.

\section{Polymer nanodroplets on chemically structured walls:
static properties}

As discussed in our previous work on droplets adsorbed on flat
walls without any chemical structuring on the substrate, the
average density distribution $\rho(R,Z)$ is a function of two
coordinates, the distance $Z$ from the substrate (note that we
orient the z-axis perpendicular to the surface and the origin of
the coordinate system is the projection of the center of mass of
the droplet on the substrate plane $z=0$), and the 
distance $R$ from the z-axis.
While individual configurations of the droplet
due to statistical fluctuations depend on the angle $\psi$
relative to the x-axis as well \cite{36}, the average density
distribution must have rotation symmetry around the z-axis
\cite{36,37,38}. When we create a chemical structure on the
surface such that the constant $\epsilon_{\textrm{w}}$
describing the wall potential $U_{\textrm{wall}}(Z)$ in
Eq.~(\ref{eq3}) has one value 
$\epsilon_{\textrm{w}}=0.2$ for $0<R<R_D$ and a smaller value
$\epsilon_{\textrm{w}}=0.05$ outside this circle whereby the symmetry
axis of the droplet coincides with the axis perpendicular to the
midpoint of this lyophilic circle, of course. Therefore it is
meaningful to record density profiles $\rho(R,Z)$ of the same type
to characterize the average droplet density profile as in our
previous work. Fig.~\ref{fig1} shows a few representative
examples. While for $R_D=15$ the droplet sits fully inside the
lyophilic circle and hence hardly differs from a droplet on an
infinitely extended lyophilic surface, for $R_D \leq 12$ the
droplet always extends over the full range of the lyophilic
region. One thus can see that the shape of the adsorbed droplet
changes when the radius $R_D$ of the lyophilic domain decreases.
As expected, the contact angle then varies continuously with
$R_D$, due to the interaction of the contact line and the boundary
between the lyophilic and lyophobic regions at the substrate, and
is no longer identical to the contact angle that applies for an
infinitely extended flat lyophilic substrate (this contact angle
is determined, for very large droplets, in terms of the
polymer-wall and polymer-gas interfacial energies, through the
Young equation \cite{23,24} while for not so large droplets also a
correction due to the line tension \cite{34,35} needs to be taken
into account \cite{36,48}). As discussed in our earlier work
\cite{36}, for nanodroplets considered here, some ambiguity in the
definition of the contact-angle for these nanodroplets is
inevitable. We follow the earlier work \cite{36,37,38}, fitting a
straight line to the density contour $\rho(Z,R)=1$ in the regime
$2 \leq Z \leq 4$, disregarding the slight curvature of this
contour in that region. Fig.~\ref{fig2} presents a plot of the
resulting variation, showing that $cos (\theta)\approx const$ for
$R_D\geq 12$, while $cos (\theta) <0$ for $R_D<9$. Comparing the
density profiles shown in the insert of Fig.~\ref{fig2} with those
of Fig.~\ref{fig1} one sees that indeed for $R_D>12$ the droplet
shape in Fig.~\ref{fig1} is the same as that for a homogeneous
surface with $\epsilon_w=0.2$. Qualitatively, the behavior seen in
Figs.~\ref{fig1},\ref{fig2} nicely corresponds to the theoretical
predictions of Lipowsky et al.\cite{12,13,14,15,16,17,18}. When
$R_D$ gets very small, the droplet shape gradually approaches the
shape that a droplet takes on a uniformly lyophobic substrate
surfaces, as the comparison of the droplet profile for $R_D=3$ in
Fig.~\ref{fig1} and the droplet profile for $\epsilon_w=0.05$ in
Fig.~\ref{fig2} shows.

We next study the behavior of a slit pore of width $L$, where both
walls exhibit a lyophilic domain of radius $R_D$ exactly opposite
to each other. When $L$ is large enough, in equilibrium (for a
fixed total number ${\mathcal{N}}$ of chains) we expect that
droplets containing ${\mathcal{N}}/2$ chains each (assuming
${\mathcal{N}}$ is an even integer) will be adsorbed exactly
opposite to each other. However, when the distance L between the
plates gets smaller, the two droplets start to interact, and a
liquid bridge between both walls can form. This formation of
liquid bridges by variation of L is illustrated in
Fig.~\ref{fig3}, choosing $R_D=14$ and ${\mathcal{N}}/2=128$, so
that the single non-bridging droplets on the separated walls are
exactly equivalent to the situation considered in
Figs.~\ref{fig1},~\ref{fig2}. One can see that for $L=22$ the
droplets are still separated and identical in shape to those seen
in Fig.~\ref{fig1} for $12 \leq R_D \leq15$. For $L \leq 19$,
however, bridge formation has occurred. One can clearly see the
change of the bridge morphology with decreasing distance
L between the two walls. When the separated droplets touch each
other, they form an ``in-bridge'' (i.e., the curvature of the
bridge surface is concave) which is typical for liquids wetting
the substrate. This catenoid shape is displayed for the profiles when the
pore width is $L=17,15,$ or $13$, respectively. On further
decrease of L, the shape of the liquid bridge becomes perfectly
cylindrical (the shapes for $L=10$ and $L=9$ are close to this
shape), while for still smaller slit pore widths $L$ (such as
$L=6$ and $L=5$) the shape is that of an ``out bridge'', with a
convex curvature of the bridge surface, i.e. a barrel-like shape.
This shape is typical for lyophobic substrates, as experienced
here by those monomers of the polymer droplet whose coordinates
$R(X,Y)$ exceed the radius $R_D$ of the hydrophilic domain. Again
the behavior in Fig.~\ref{fig3} is very nicely consistent with the
behavior as predicted by the theory for quasi-macroscopic droplets
\cite{12,13,14,15,16,17,18,18a}, and hence we again find that these
phenomenological concepts due to Lipowsky et al.
\cite{12,13,14,15,16,17,18} work qualitatively down to the
nanoscopic scale.

In view of the fact that there is some ambiguity in the precise
numerical estimation of the contact angle of nanodroplets, as
mentioned above, it is of significant interest to estimate
additional properties quantitatively, which could be compared to
analytical theories on this problem. Such properties are the base
radius $R_{\textrm{lat}}$ of the droplet (Fig.\ref{fig4}) or
liquid bridge (Fig.\ref{fig5}), the height of the droplet $H$
(Fig.\ref{fig4}), and the midheight radius $R_{\textrm{lat}}(Z=H/2)$
of the droplet (Fig.~\ref{fig4}) or of a liquid bridge
$R_{\textrm{lat}}(Z=L/2)$, (Fig.~\ref{fig5}), respectively.
Figs.~\ref{fig4},\ref{fig5} demonstrate that these quantities can
be measured with relatively small statistical errors in our
simulations. From such data it also is evident that a pronounced
change in the behavior occurs at $R_D\approx 8.5$
(Fig.~\ref{fig4}). For a macroscopic sessile droplet (satisfying
the Young equation with the contact angle $\theta$ and having a
sphere-cap shape) we simply would have the relations in terms of
the sphere radius $r$
\begin{equation}\label{eq4}
R_{\textrm{lat}}=  r\;\sin (\theta), \; H=r(1-\cos
(\theta))\;,
\end{equation}
and hence
\begin{equation}\label{eq5}
H/R_{\textrm{lat}}=(1-cos (\theta))/\; \sin \;(\theta) \approx
\theta /2\;, \quad \theta \rightarrow 0\;.
\end{equation}
Indeed this relation is roughly fulfilled for our model.

It also is of interest to calculate the pressure in the liquid
bridge, using the virial formula \cite{45,49}. Fig.(~\ref{fig6})
shows that the pressure is positive for the smallest distances
between the plates $(L \leq 6)$ only, while for larger distances
the pressure is negative. This observation already indicates that
this situation is unfavorable, if we would not enforce the
distance $L$ between the plates as given parameter, such distances
would not occur if the walls could freely move against each other.
This fact is very clearly borne out when we compute the normal
force between the walls (Fig.~\ref{fig7}). Of course, for large
distances $L$ for which two separate droplets occur on each wall
that do not yet touch the force is zero, while for bridging
droplets an attractive force arises, and only for very small $L$,
where the bridge is squeezed into the lyophobic part, the force
becomes repulsive (Fig.~\ref{fig7}). The force goes through zero
at $L \approx 6$. Since it has been shown\cite{18a} that for the
underlying {\em catenoid} geometry a closed analytical expression
for the surface area dependence on the wall separation $L$ does not exist,
we tentatively use a relation proposed by Swain and Lipowsky\cite{14}
for the bridge between a single pair of opposing lyophilic stripes. 
The corresponding contact angle $\theta ^*$ can be calculated then
from $\theta ^* = - tan ^{-1} (2R_D/L) \approx 100^o$ and agrees
with our observations. As
expected, this contact angle exceeds $\pi/2$. The strongest
attractive force actually does occur near $L\approx 10$, where the
angle is $\pi/2$. For $L > 12$ an almost linear decrease of the
force sets in, before it discontinuously jumps to zero for $L=22$.

It is interesting to note that very long range attractive forces
have been experimentally detected in AFM measurements of forces
between a polystyrene sphere and liquid interfaces \cite{21,22},
showing a {\em linear} variation with distance on the scale of 30 nm.
Unlike such experiments, in our simulation an electrostatic origin
of such long range forces is excluded by construction of our
model. The force seen in in Fig.~\ref{fig7} is entirely due to the
interplay of the various interfacial interactions that control the
shape of the liquid bridge. One should also point out that both
the course of the bridging force against inter-plate distance $L$ as 
well as the particular bridge configurations corresponding to various
parts of the force distance curve closely match some recent
results\cite{18a} on classical structureless droplets obtained by two 
different surface minimization techniques.

\section{Dynamic aspects of bridge formation}

For $L=21.5$ one can observe initial states where the two substrates
still carry separate droplets, but the wings of their density
distributions already overlap, and this interaction between the
two droplets starts a merging process which leads to the formation
of a liquid bridge. Figs.(~\ref{fig8}) and (\ref{fig9}) analyze
the dynamical aspects of this merging process of the two droplets
in more detail. One can see that the formation of the liquid
bridge is a very slow process, it is clearly diffusion-controlled,
there is no evidence of faster hydrodynamic mechanisms. The radius
of the midpoint of the bridge seems to grow towards its
equilibrium value over a transient period of time according to a
$t^{1/4}$ law. This behavior is reminiscent of the growth law with
which an interfacial profile between coexisting phases approaches
equilibrium \cite{50}.

\section{Concluding discussion}

In the present work, we have presented Molecular Dynamics
simulations addressing the shape of a droplet adsorbed on a
spherical lyophilic domain on an otherwise lyophobic flat
substrate surface, and the formation of liquid bridges between two
such surfaces. Also the forces between these surfaces caused by
such liquid bridges have been measured, and the kinetics of the
merging of two such droplets into one bridge has been studied.

\begin{figure}[h]
\newpage
.
\begin{figure}
\includegraphics{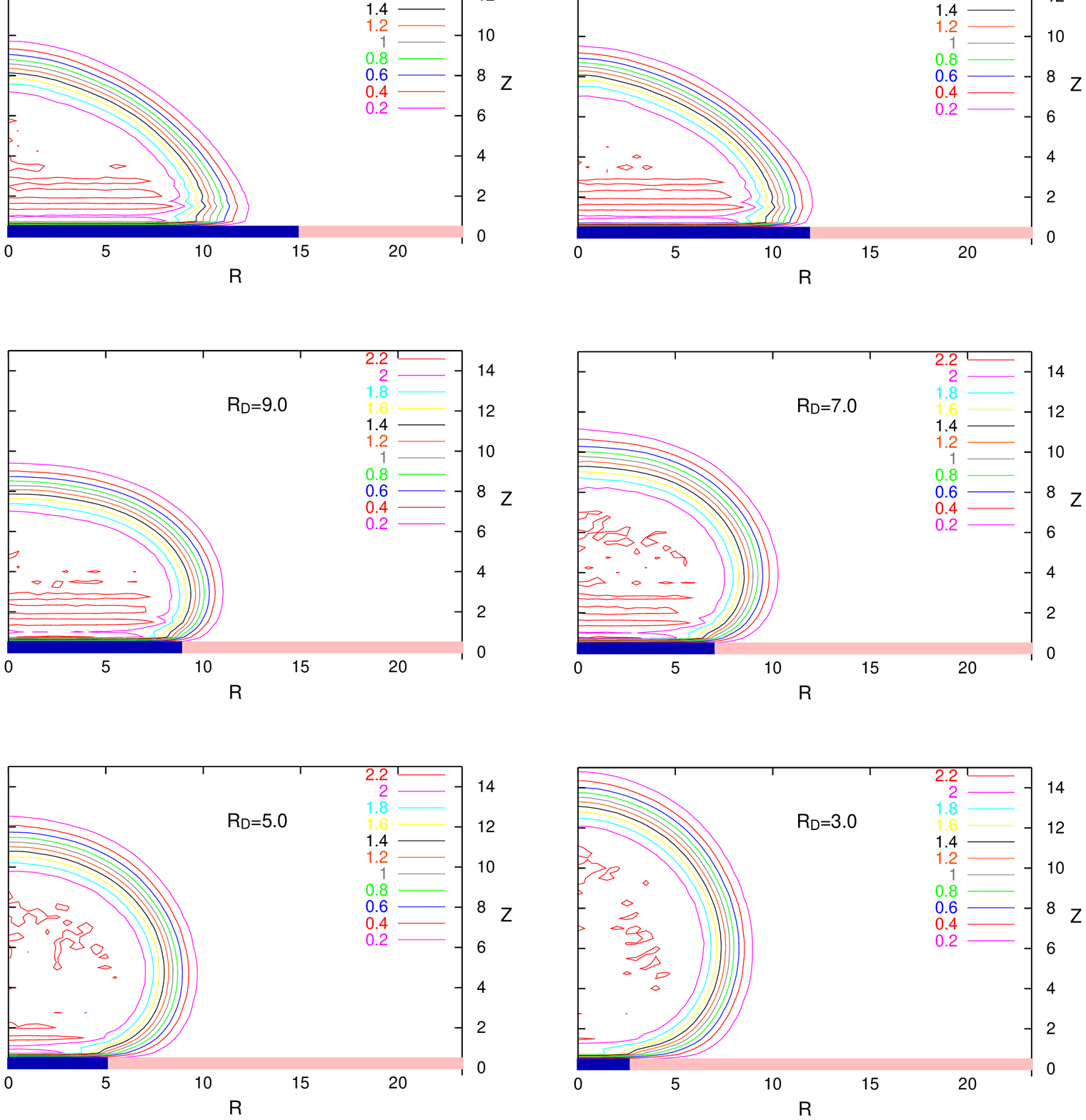} 
\vskip 18.5cm
\caption{\label{fig1}}
\end{figure}

\newpage
.
\begin{figure}
\includegraphics{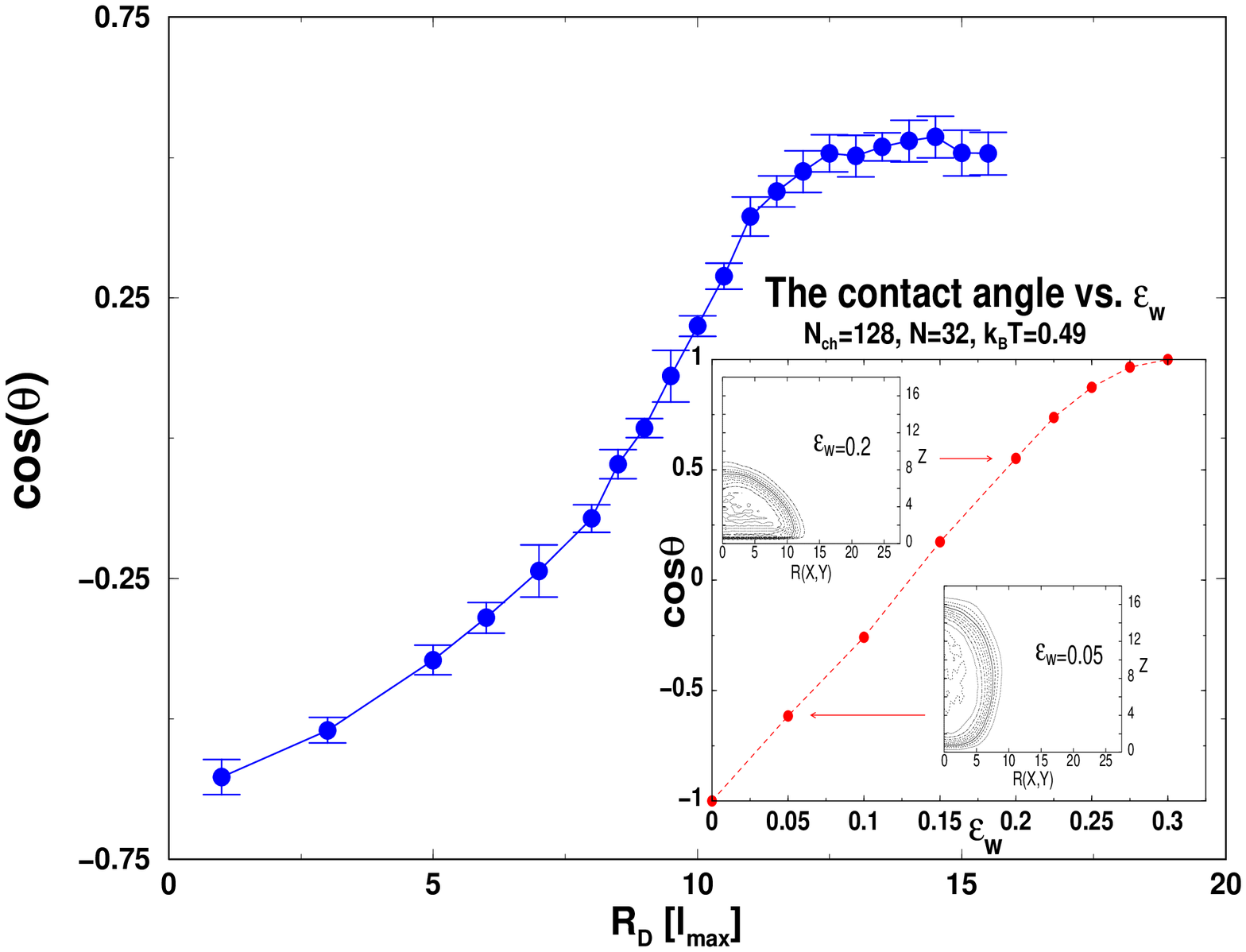} 
\vskip 24.3cm
\caption{\label{fig2}}
\end{figure}

\newpage
.
\begin{figure}
\includegraphics{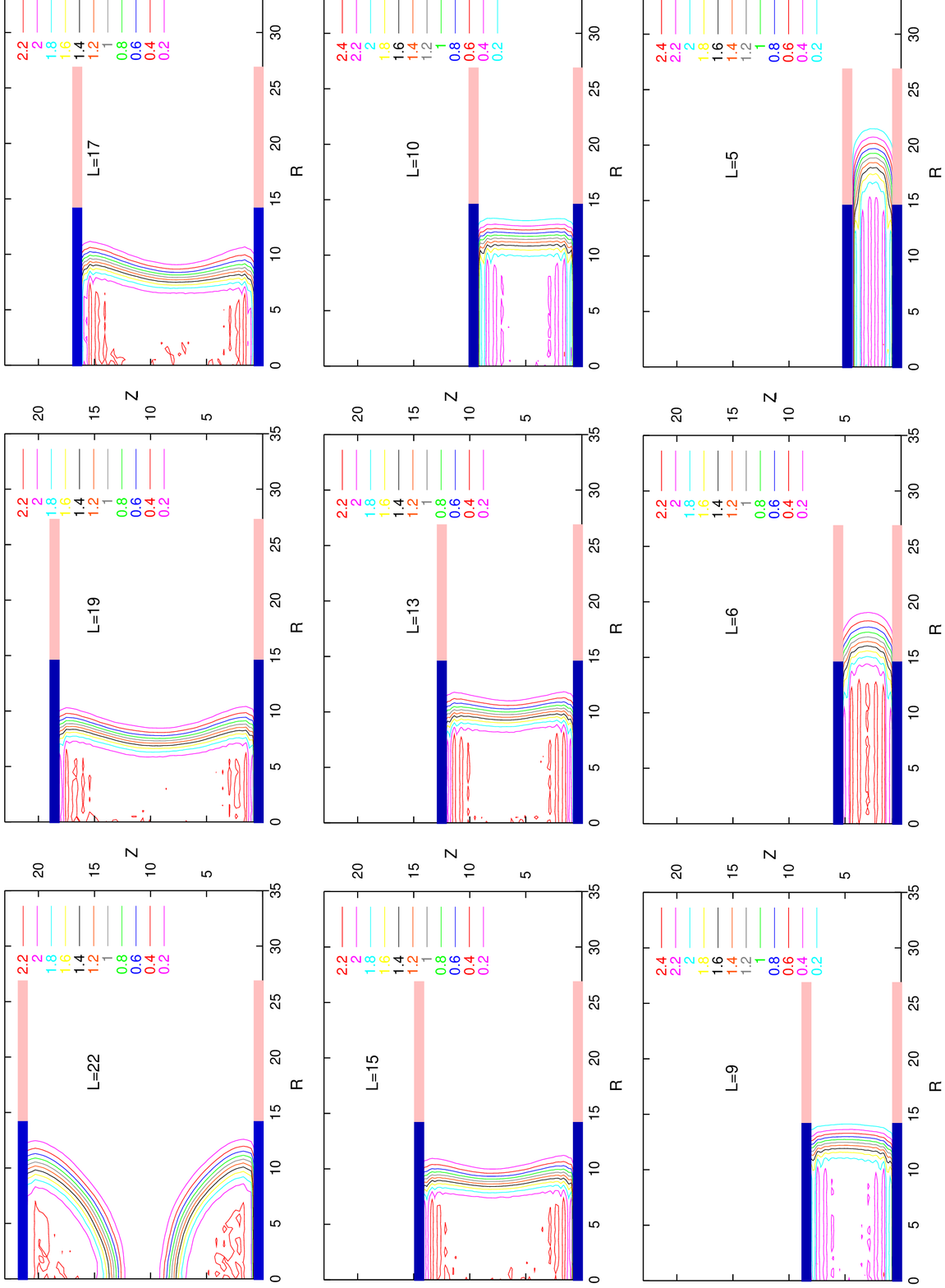} 
\vskip 23.5cm
\caption{\label{fig3}} 
\end{figure}

\newpage
.
\begin{figure}
\includegraphics{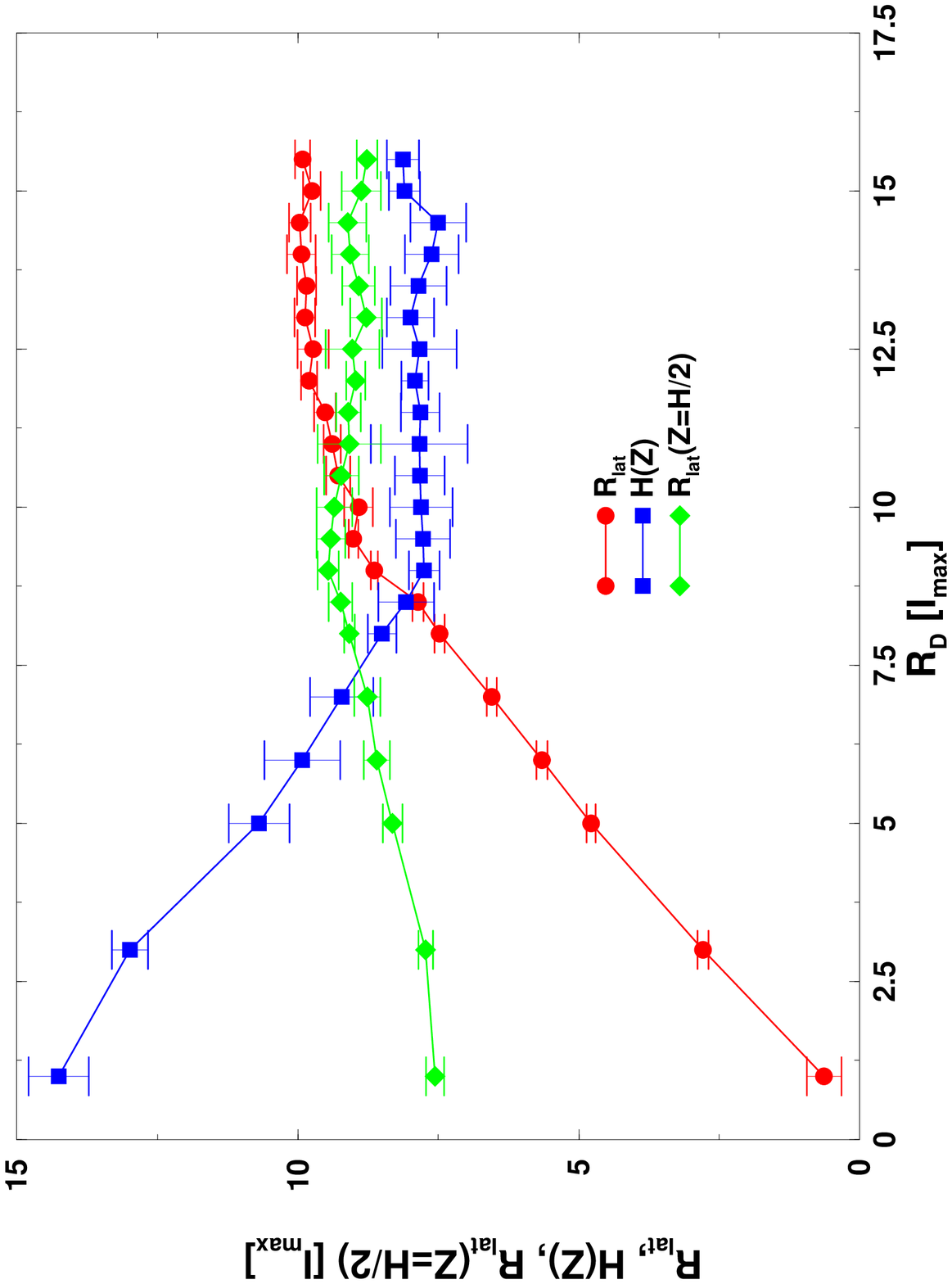} 
\vskip 17.5cm
\caption{\label{fig4}} 
\end{figure}

\newpage
.
\begin{figure}
\includegraphics{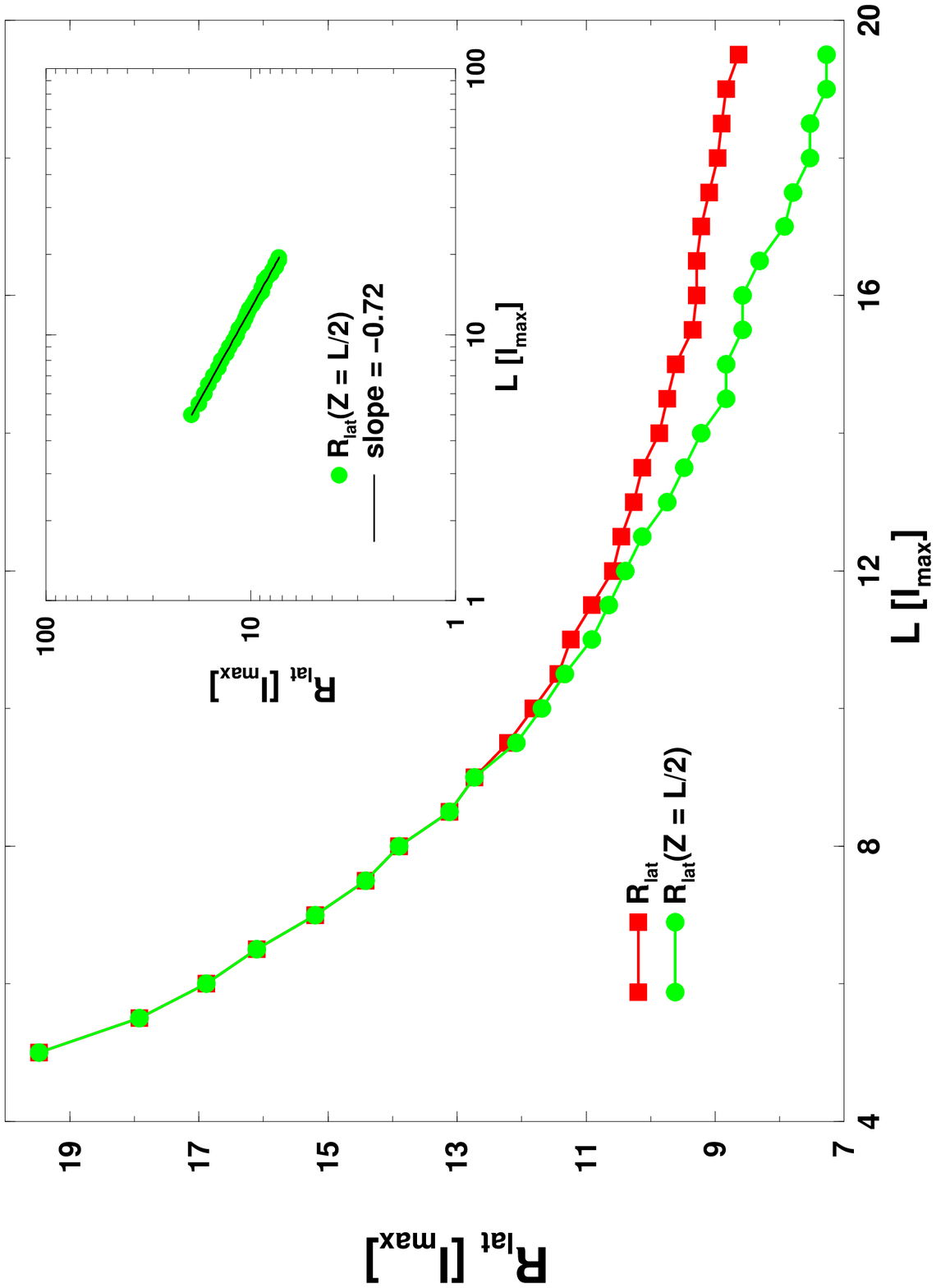} 
\vskip 17.5cm
\caption{\label{fig5}}
\end{figure}

\newpage
.
\begin{figure}
\includegraphics{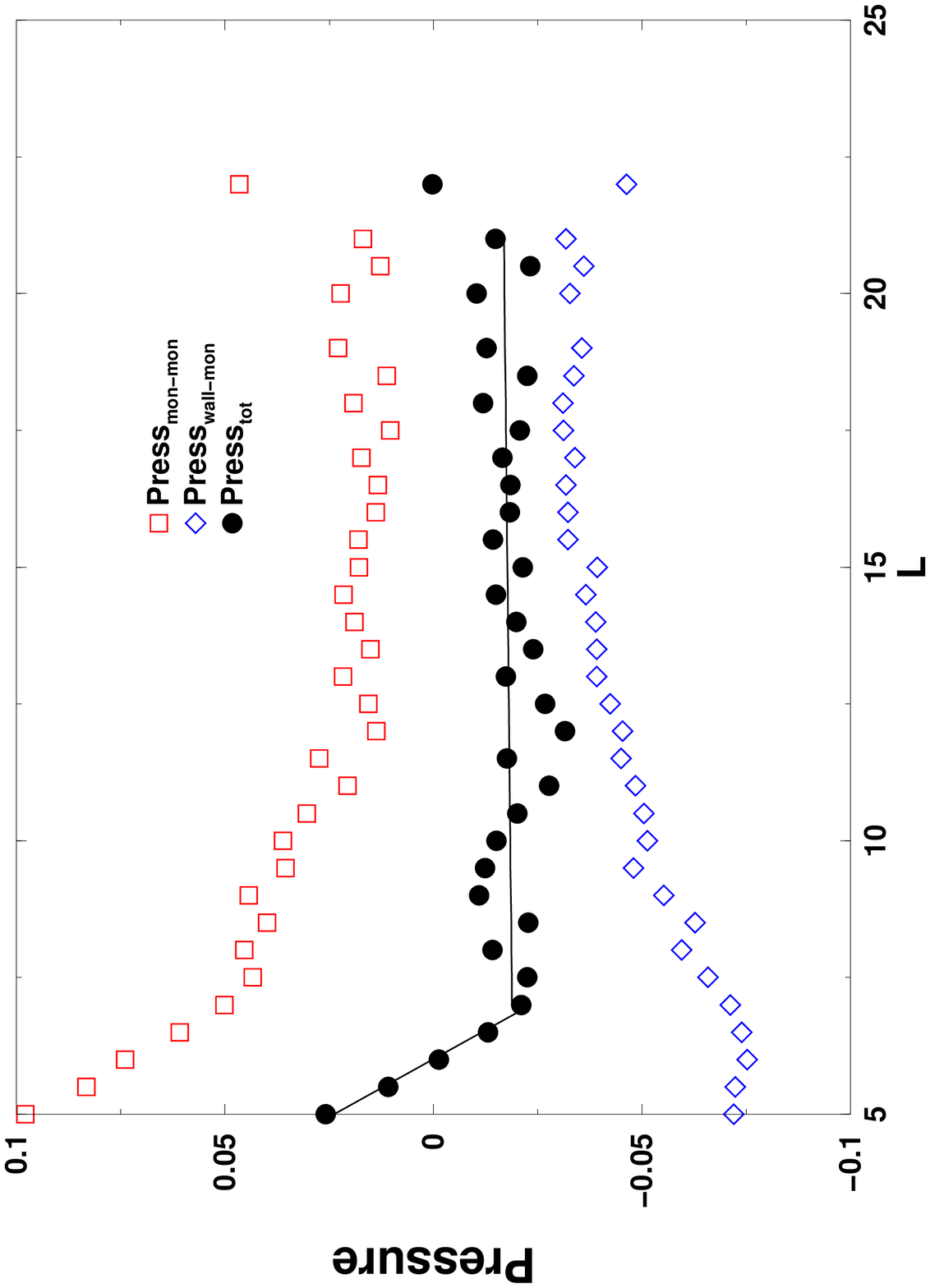} 
\vskip 17.5cm
\caption{\label{fig6}} 
\end{figure}

\newpage
.
\begin{figure}
\includegraphics{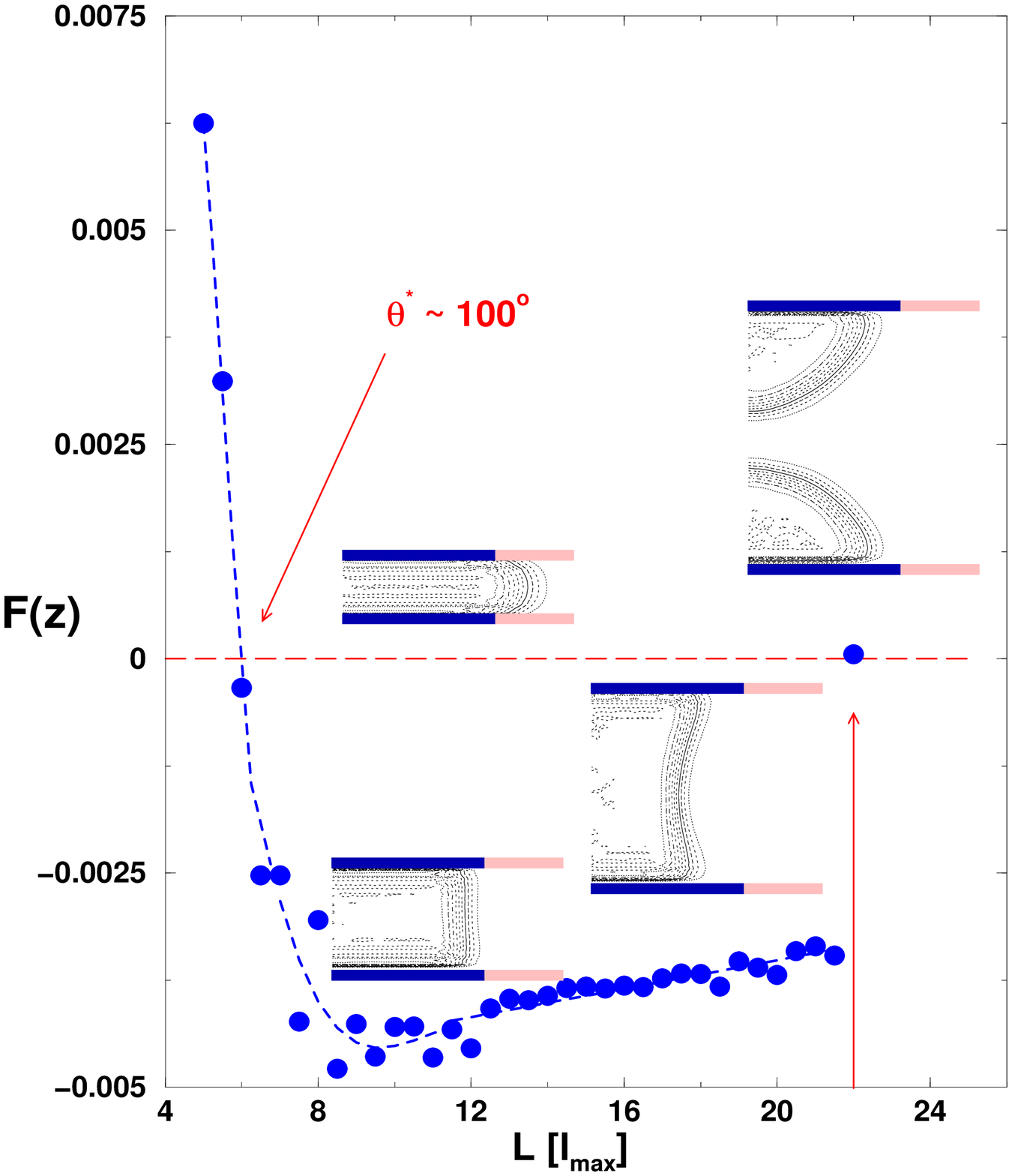} 
\vskip 22.5cm
\caption{\label{fig7}} 
\end{figure}

\newpage
.
\begin{figure}
\includegraphics{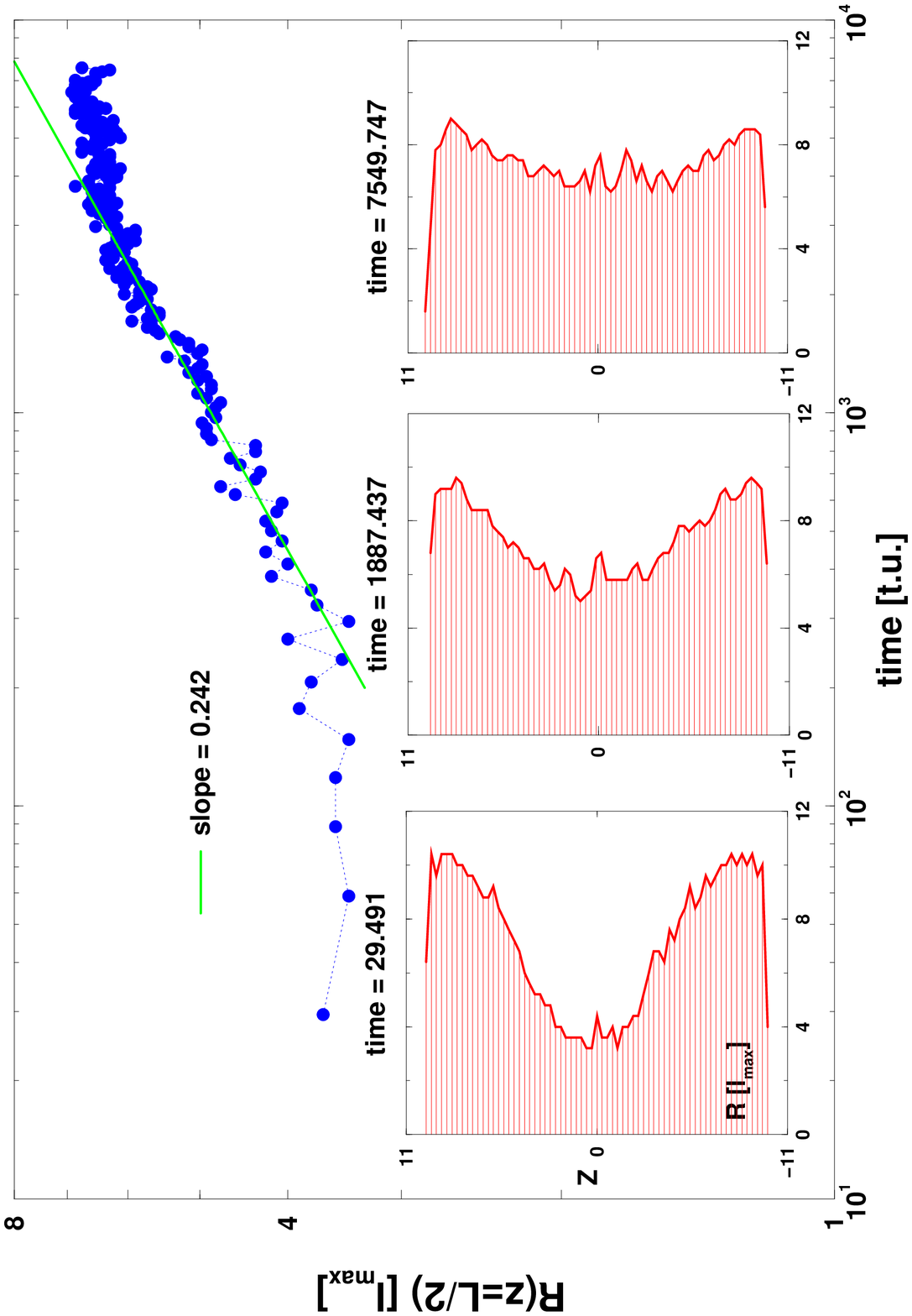} 
\vskip 17.5cm
\caption{\label{fig8}}
\end{figure}

\newpage
.
\begin{figure}
\includegraphics{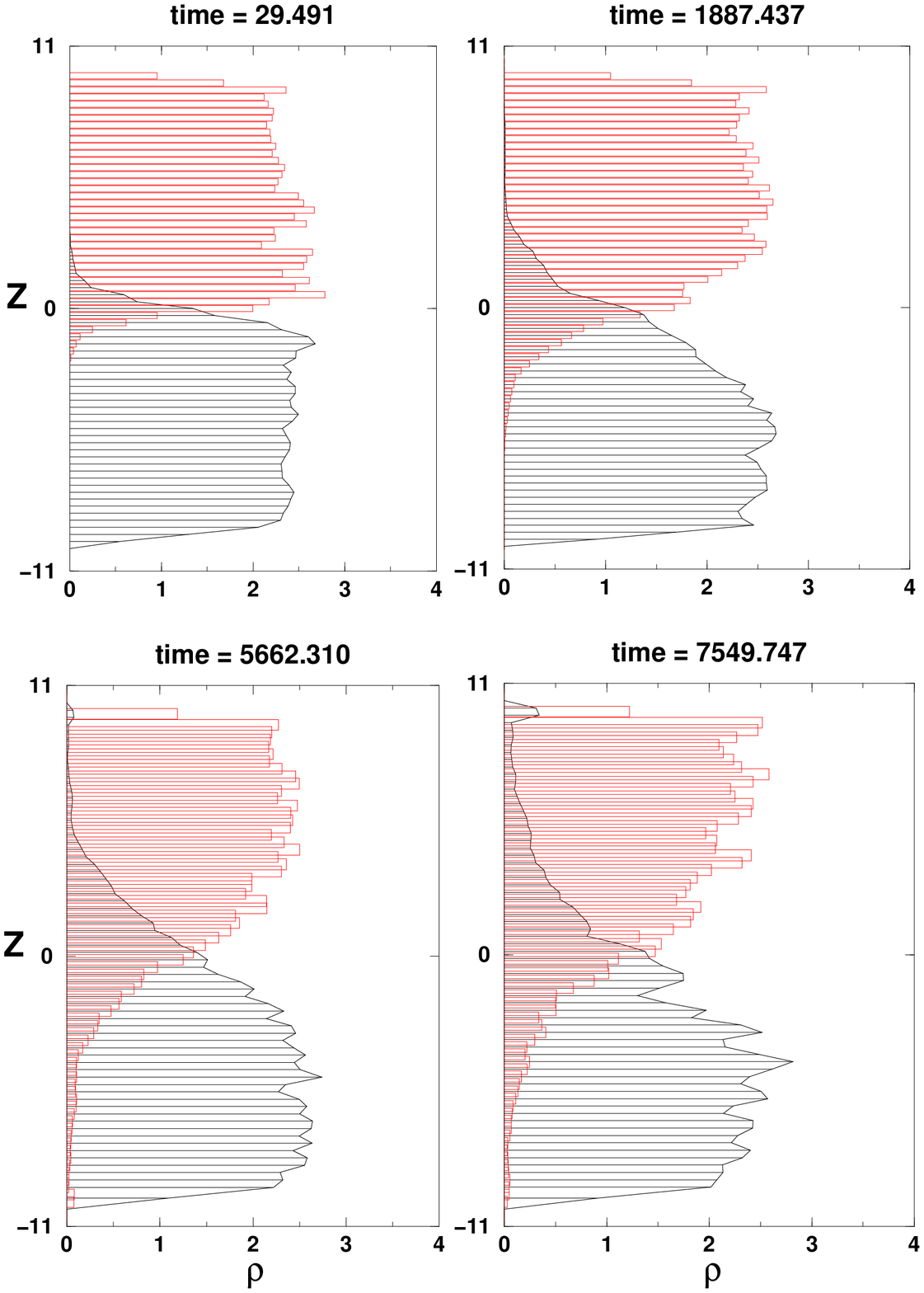} 
\vskip 19.5cm
\caption{\label{fig9}}
\end{figure}

\end{figure}

According to the predictions of Lipowsky et al.
\cite{12,13,14,15,16,17,18} one should expect for large enough
domain radius $R_D$ that the contact angle $\theta$ of the droplet
is constant ($\theta= \theta_\textrm{phil}$, the ``lyophilic''
value, with $\cos \theta_\textrm{phil} \approx 0.57$ in our case,
see inset of Fig.~\ref{fig2}), until the radius $R_\textrm{lat}$
of the droplet matches $R_D$. For our choice of parameters, this
happens for $R_D = R_D^\textrm{phil} \approx 12$. For $R_D <
R_D^\textrm{phil}$, one expects that $\theta$ should increase such
that $R_\textrm{lat}=R_D$ is always maintained, until $\theta$
reaches the value of the lyophobic part of the substrate surface,
$\theta=\theta_\textrm{phob}$, for $R_D=R_D^\textrm{phob}$, with
$\cos \theta_\textrm{phob} \approx - 0.6$ in our case. In fact,
Fig.~\ref{fig4} nicely verifies the linear variation
$R_\textrm{lat}=R_D$ quantitatively up to about $R_D \approx 9$,
while for $9 < R_D < R_D^{\textrm{phil}}$ the further increase of
$R_\textrm{lat}$ with $R_D$ is slower, and the saturation value of
$R_\textrm{lat}$, reached for $R \geq R_D^{\textrm{phil}}$, is in fact
smaller than expected, namely only around
$R_\textrm{lat}^\textrm{max}\approx 10$. Of course, some
quantitative deviations of our results from the predictions of
Lipowsky et al. \cite{12,13,14,15,16,17,18} must be expected,
since the latter predictions are asymptotically valid for very
large, almost macroscopic, droplets, while our simulations concern
nanodroplets, and hence there are no sharp transitions possible
when $R_D$ is varied: so Fig.~\ref{fig2} does not show a sharp
kink of the curve $\cos (\theta) $ vs. $R_D$ at $R_D =
R^{\textrm{phil}}_D$ but rather a rounded crossover, and also at
$R_D = R_D^{\textrm{phob}}$ only a very smooth variation is seen
(in fact, $R_D^\textrm{phob}$ is of order unity in our case, and
hence cannot be even uniquely identified from our data). Clearly,
it would be interesting to analyze the amount of rounding of these
``transitions'' at $\theta=\theta_\textrm{phil}$ and
$\theta=\theta_\textrm{phob}$ theoretically, and to repeat our
simulations for larger droplets (and correspondingly chosen larger
values of $R_D$) to see the extent to which these transitions
become sharper, but all such extensions of our work would present
major difficulties and hence have not been attempted.

We also note that the theory implies that the shape of the droplet
should be a sphere cap throughout. In particular, since the volume
$V$ of the droplet is constant, we should have

\begin{equation} \label{eq6}
V=\frac{\pi H}{6} (3 R^2_\textrm{lat} + H^2) = \textrm{const}
\end{equation}

and combining this equation with the geometrical relations for the
contact angle , Eqs.~(\ref{eq4}),~(\ref{eq5}), in principle the
variation of $\cos (\theta)$ with $R_D$ in the regime
$R^\textrm{phob}_D < R_D < R_D^\textrm{phil}$ can be explicitly
predicted. Noting from Fig.~\ref{fig4} that for $R_D >
R_D^\textrm{phil}$ we have $R_\textrm{lat}\approx10$, $H \approx
7.5$, one obtains $V \approx 1400.$ However, for $R_D \approx 7$
where $R_\textrm{lat} \approx6.8$, $H \approx9.3$
(Fig.~\ref{fig4}) we find that Eq.~(\ref{eq6}) would yield only $V
\approx 1100$, and for $R_D \approx5$, where
$R_\textrm{lat}\approx5$, $H \approx11$ we would get $V \approx
1130$, while for $R_D=R_\textrm{lat}=3$ where $H\approx13$ we find
$V \approx 1330$. Thus, the geometrical relations on the basis of
the sphere cap picture are not verified accurately in our case.
This problem is related to the fact that the linear increase of
$R_\textrm{lat}$ with $R_D$ does not continue all the way to
$R_\textrm{lat}=R_D^\textrm{phil}\approx12$ but $R_\textrm{lat}$
is significantly too small for $R_D >9$. On the other hand, the
discrepancies found are not dramatic either, they do not exceed
10-15\%, in spite of the nanoscopic size of the droplets. One
needs also to consider the fact that this small size leads also to
substantial errors and ambiguities when one tries to read off
$R_\textrm{lat}$ and $H$ from the data. In view of all these
problems, the agreement between the simulations and the
theoretical descriptions certainly is satisfactory.

\bigskip
\underline{Acknowledgments}: This research was supported in part
by the Deutsche Forschungsgemeinschaft (DFG) under grant number
436 BUL 113/130/2-1.

\clearpage

\newpage

FIG. 1. Contour diagrams representing the density
profiles $\rho(Z,R(X,Y))=\nu \Delta \rho$, $\nu =1,2,...11$,
$\Delta \rho =0.2$, in the $(Z,R)$ plane, for a droplet containing
128 chains with 32 monomers each, at a temperature $k_BT=0.49$.
The droplet is in contact with an ideally flat lyophobic substrate
(represented by the thick grey line in the bottom) decorated with
one lyophilic circle of radius $R_D$ (the thick black line in the
bottom). The adsorption strengths of the lyophobic wall is
$\epsilon _{\textrm{w}}=0.05$, while for the lyophilic circle
it is $\epsilon_{\textrm{w}}=0.2$. Profiles are shown for
$R_D=15,12,9,7,5$ and $3$, respectively. These profiles are
obtained by averaging over 10 runs of $2.1 \cdot 10^6$ integration
steps each. One MD time step is $\delta t=0.0009$ MD time units.\\[1cm]
FIG. 2. Cosine of the contact angle
$\theta$ plotted versus the radius $R_D$ of the lyophilic domain,
for droplets containing ${\mathcal{N}}=128$ chains with $N=32$
monomers each, at $k_BT=0.49$. The inset shows the dependence of
the contact angle of the droplet at an infinitely extended
homogeneous substrate (as considered in Refs.\cite{36,37,38}) on
the strength $\epsilon_{\textrm{w}}$ of the wall potential. The
two contour diagrams in the inset represent the density profiles
$\rho(Z,R(X,Y))=\nu\Delta\rho ,\; \nu =1,2,\ldots,11$ in the
$(Z,R)$ plane of a droplet for the two different adsorption
strengths relevant for the present paper, namely for
$\epsilon_{\textrm{w}}=0.05$ (the value used to model the
lyophobic region of the wall) and for
$\epsilon_{\textrm{w}}=0.2$ (the value used for the lyophilic
domain on the surface).\\[1cm]
FIG. 3. Contour diagrams representing
the density profiles $\rho(Z,R(X,Y))=\nu \Delta \rho,\; \nu
=1,2,\ldots,11 (\Delta \rho =0.2)$ in the $(Z,R)$ plane for two
droplets adsorbed on opposite walls $(L=22)$ or a liquid bridge
between the walls $(L=19,17,15,13,10,9,6,5)$ respectively. The
simulated system contains 256 chains with 32 monomers each, the
temperature is chosen as $k_BT=0.49$, and the walls are ideally
flat and lyophobic, with an adsorption strength $\epsilon_w=0.05$,
except for a lyophilic circle of radius $R_D=14$ (with adsorption
strength $\epsilon_w=0.2$). The lyophilic parts of the adsorbing
surface are shown by thick black bars, the lyophobic parts by
thick grey bars. The profiles are averaged over 18 runs of $1.05
\cdot 10^6$ MD steps.\\[1cm]
FIG. 4. Variation of the base radius
$R_{\textrm{lat}}$ (circles), the height H (squares) and the
midheight radius $R_{\textrm{lat}}(Z=H/2)$ (diamonds) of a droplet
adsorbed on a single circular lyophilic domain of radius $R_D$
plotted vs. $R_D$ at $k_BT=0.49$. The droplet contains 128 chains
with 32 monomers each. The strength of the adsorption potential is
$\epsilon_{\textrm{w}}=0.2$ inside the lyophilic domain and
$\epsilon_{\textrm{w}}=0.05$ outside of it. The radii and the
height are measured from density profile contour diagrams (taking
the contour at midpoint density, $\rho=1$) averaged over 10 runs
of $2.1 \cdot 10^6$ integration steps each. $R_{\textrm{lat}}$ is
taken as the maximum lateral extension of the contour (it occurs
roughly at $Z=1.5$). The error bars are obtained by estimating the
radii from every single run.\\[1cm]
FIG. 5. Variation of the base radius
$R_{\textrm{lat}}$ (circles) and the midheight radius
$R_{\textrm{lat}}(Z=L/2)$ (squares) for the liquid bridges of
Fig.~\ref{fig3} (for further
explanations see the caption of Fig.~\ref{fig4}).
A log-log plot of $R_{\textrm{lat}}(Z=L/2)\; \mbox{vs}\; L$  (see inset) reveals a
power-law relationship $R_{\textrm{lat}}\propto L^{-0.72}$.\\[1cm]
FIG. 6. The pressure in a liquid
bridge connecting two flat substrates is plotted against the
distance $L$ between the walls. The total pressure is shown by
filled circles. Open diamonds show the contribution to the
pressure that comes from the wall-monomer interactions, while the
open squares show the contribution to the pressure from the
monomer-monomer interaction and the kinetic part.\\[1cm]
FIG. 7. The normal force acting
between two substrates with a bridging droplet plotted versus the
distance $L$. The insets show contour diagrams of the density
profiles of the liquid bridge at 4 different distances $L$: the
two separated droplets are at $L=22$, the ``in-bridge'' is at
$L=17$, the cylindrical bridge is at $L=9,5$ and the
``out-bridge'' at $L=6$. The thick dark lines indicate the
lyophilic region of the substrate. Each point is obtained from
averaging over 19 runs of $1.05 \cdot 10^6$ MD steps, while the
data points around the minimum ($6.5 < L < 8.5$) are averaged over
29 runs. The arrow at $L=22$ indicates that bridges are no longer
stable for $L \geq 22$. The contact angle at zero force is $\theta ^* 
\approx 100^0$.\\[1cm]
FIG. 8. Midheight radius $R(Z=L/2,t)$
of a forming liquid bridge as a function of time, for $L=21.5$. The
radial distance from the vertical axis of the droplet to the point
where the density is $\rho = 1.0$ is measured. The origin of time
$t=0$ is chosen as the time when the two droplets sitting on
opposite parallel substrates have just touched each other. The
insets show vertical cross sections of the resulting
bridges for three different times indicated in the figure.
$R(Z=L/2,t)$ is averaged over 7 independent runs. For $300 \leq t
\leq 3000$ an effective growth law $R(Z=L/2,t)\propto const +
t^{0.242}$ is observed, as indicated by a straight line.\\[1cm]
FIG. 9. Density profiles of a forming
liquid bridge for four different times, as indicated in the
figure. The diffusive interpenetration of species which originally
belong to the two separated droplets is indicated.

\end{document}